\documentclass[
  prx,
  10pt,
  reprint,
  eprint,
  twocolumn,
  raggedbottom,  % this goes with the twocolumn option
  amsmath,
  amsfonts,
  amssymb,
  dvipsnames,
  superscriptaddress,
]{revtex4-2}

\usepackage[utf8]{inputenc}
\usepackage{microtype}
\usepackage{xparse}

\usepackage{booktabs}
\usepackage{dcolumn}
\usepackage[hidelinks]{hyperref}
\usepackage[all]{hypcap}
\usepackage{siunitx}
\usepackage{bm}
\usepackage[ISO]{diffcoeff}
\usepackage[normalem]{ulem}

\usepackage{graphicx}
\usepackage{xcolor}
\usepackage{newtxtext}
\usepackage[varvw]{newtxmath}

\newcommand*{\dd}{\text{d}}

\newcommand*{\gradient}{\bm{\nabla}}

\newcommand*{\laplacian}{\nabla^2}

\newcommand*{\rms}{\text{rms}}
\newcommand*{\Emean}[1]{\langle #1 \rangle}  % Eulerian mean

\newcommand*{\xvec}{\bm{x}}
\newcommand*{\kvec}{\bm{k}}
\newcommand*{\rvec}{\bm{r}}

\newcommand*{\Lforcing}{L_\text{f}}

\newcommand*{\vvec}{\bm{v}}

\newcommand*{\vort}{\omega}
\newcommand*{\vortvec}{\bm{\vort}}

\newcommand*{\Rey}{\text{Re}}     % Reynolds number
\newcommand*{\diss}{\varepsilon}  % energy dissipation
\newcommand*{\lambdaTaylor}{\lambda}
\newcommand*{\Rlambda}{R_{\lambdaTaylor}}
\newcommand*{\vrms}{v_\rms}

\newcommand*{\epsinj}{\widetilde \diss_{\text{inj}}}
\newcommand*{\epsinjn}{\diss_{\text{inj}}^\text{n}}
\newcommand*{\epsinjs}{\diss_{\text{inj}}^\text{s}}
\newcommand*{\epsvisc}{\widetilde \diss_{\nu}}
\newcommand*{\epsMF}{\diss_{\text{ns}}}
\newcommand*{\epsN}{\diss_{\text{n}}}
\newcommand*{\epsS}{\diss_{\text{s}}}
\newcommand*{\kmax}{k_{\text{max}}}
\newcommand*{\vortS}{\vortvec_{\text{s}}}
\newcommand*{\vortN}{\vortvec_{\text{n}}}

\newcommand*{\Stilde}{\widetilde{S}}
\newcommand*{\SStilde}{\widetilde{S}_2}
\newcommand*{\SSStilde}{\widetilde{S}_3}
\newcommand*{\nutilde}{\widetilde{\nu}}
\newcommand*{\etatilde}{\widetilde{\eta}}
\newcommand*{\lambdatilde}{\tilde{\lambda}}
\newcommand*{\nuMF}{\nu_\text{mf}}
\newcommand*{\RlambdaTilde}{R_{\lambdatilde}}

\newcommand*{\nutot}{\nu_{\text{tot}}}

\newcommand*{\dvvec}{\delta \vvec}
\newcommand*{\dvpara}{\delta v_{\parallel}}

\newcommand*{\Flatness}{\mathcal{F}}

\newcommand*{\HeFour}{$^4\text{He}$}
\newcommand*{\vn}{\vvec_{\text{n}}}
\newcommand*{\vs}{\vvec_{\text{s}}}
\newcommand*{\vns}{\vvec_{\text{ns}}}
\newcommand*{\pn}{p_{\text{n}}}
\newcommand*{\ps}{p_{\text{s}}}
\newcommand*{\nun}{\nu_{\text{n}}}
\newcommand*{\nus}{\nu_{\text{s}}}
\newcommand*{\rhon}{\rho_{\text{n}}}
\newcommand*{\rhos}{\rho_{\text{s}}}
\newcommand*{\Fns}{\bm{F}_{\text{ns}}}

\newcommand*{\OmegaS}{\Omega_{\text{s}}}
\newcommand*{\phiNS}{\varphi_{\text{ns}}}
\newcommand*{\Cns}{C_{\text{ns}}}
\newcommand*{\VLD}{\mathcal{L}}  % vortex line density

\newcommand*{\FluxT}{\mathcal{T}}
\newcommand*{\FluxTn}{\FluxT^{\text{n}}}
\newcommand*{\FluxTs}{\FluxT^{\text{s}}}

\newcommand*{\CaseTlowBis}{$\times$}
\newcommand*{\CaseTlowTer}{$\triangledown$}
\newcommand*{\CaseTmidExt}{$\circ$}
\newcommand*{\CaseTmidExtBis}{$\square$}
\newcommand*{\CaseTmidExtTer}{$\triangleright$}
\newcommand*{\CaseThighAdapted}{$\triangle$}
\newcommand*{\CaseNS}{$\blacklozenge$}

\newcommand*{\HVBK}{HVBK}

\definecolor{forestgreen}{rgb}{0.13, 0.55, 0.13}

\begin{document}

\newcommand*{\LMFA}{%
  CNRS, Ecole Centrale de Lyon, INSA Lyon, Université Claude
  Bernard Lyon 1, Laboratoire de Mécanique des Fluides et d'Acoustique, UMR
  5509, 69130 Ecully, France
}
\newcommand*{\LEGI}{%
  Univ.\ Grenoble Alpes, CNRS, Grenoble INP, LEGI, 38000 Grenoble, France
}

\newcommand*{\Neel}{%
  Institut Néel, CNRS, Univ.\ Grenoble Alpes, Grenoble F-38042, France
}
\newcommand*{\MtwoC}{%
  University of Rouen Normandy, CNRS, M2C, 76000 Rouen, France
}

\title{Disentangling temperature and Reynolds number effects in quantum turbulence}
\date{\today}

% Use letters for affiliations, numbers to show equal authorship (if applicable) and to indicate the corresponding author
\author{Juan Ignacio Polanco}
\email{juan-ignacio.polanco@cnrs.fr}
\affiliation{\LMFA}
\affiliation{\LEGI}
\author{Philippe-E. Roche}
\affiliation{\Neel}
\author{Luminita Danaila}
\affiliation{\MtwoC}
\author{Emmanuel Lévêque}
\affiliation{\LMFA}


\begin{abstract}
The interplay between viscous and frictional dissipation is key to understanding quantum turbulence dynamics in superfluid $^4$He. Based on a coarse-grained two-fluid description, an original scale-by-scale energy budget that identifies each scale's contribution to energy dissipation is derived. 
Using the Hall-Vinen-Bekharevich-Khalatnikov (\HVBK{}) model to further characterize mutual friction,
direct numerical simulations at temperatures {$1.44\,\mathrm{K} \lesssim T \lesssim 2.16\,\mathrm{K}$} indicate that mutual friction promotes intense momentum exchanges between the two fluids to maintain a joint energy cascade despite their viscosity mismatch.
However, the resulting {overall} frictional dissipation remains small (compared to the viscous dissipation) and confined to far-dissipative scales.
This remarkable feature allows us to define an effective Reynolds number for the 
turbulence intensity in a
two-fluid system, helping to disentangle the effects of Reynolds number and temperature in quantum turbulence.
Thereby, simple physical arguments predict that the distance $\ell$ between quantized vortices (normalized by the turbulence integral scale $L_0$) should behave as $\ell/L_0 \approx 0.5 \, \Rey_\kappa^{-3/4}$ with the Reynolds number based on the quantum of circulation $\kappa$. This law is well supported by a large set of experimental and numerical data within the temperature range of the \HVBK{} model.
Finally,
this new approach offers the possibility of revisiting the ongoing controversy on intermittency in quantum turbulence.
It is shown that observed changes in intermittency arise from Reynolds number effects rather than from temperature variations, as proposed in recent studies.
\end{abstract}

\maketitle

The superfluidity of liquid \HeFour{} below $T_\lambda=2.1768\,\mathrm{K}$ (referred to as He-II) is a remarkable manifestation of quantum effects at {hydrodynamic} scales \cite{donnelly2005quantized}.
The {two-fluid model} initiated by Tisza {\cite{Tisza1938,Tisza1940a,Tisza1940b,BALIBAR2017586}} and further developed by Landau \cite{Landau41} describes He-II as a mixture of two co-penetrating fluids, each with its distinct velocity: a viscous {normal  fluid} ($\vn$),
and a {superfluid} ($\vs$) that moves without viscosity and has rotational motions confined to vortex lines with quantized circulation $\kappa  \approx  0.997~10^{-7}~\text{m}^2/\text{s}$ \cite{Landau41}.
In turbulent He-II flows, these quantum vortices are primarily responsible for a mutual friction~\cite{Vinen2002} that couples the two fluids, acting as a resistance to their {relative} motion.
To complete this picture, the mass density of each fluid strongly varies with temperature, thus making the strength of mutual friction temperature-dependent \cite{Barenghi2014a}. 
In the context of He-II turbulence, also known as quantum turbulence~\cite{Skrbek2021,Barenghi2023}, mutual friction plays a role in the energy cascade, alongside the viscous dissipation of the normal fluid, which is also present in classical fluid turbulence.
A central question is to determine at which scales, and to what extent, mutual friction affects scaling properties~\cite{Sreenivasan1997} and how this varies with temperature. 
 
Generally speaking, mutual friction tends to reduce the slip velocity $\vn-\vs$, with the component having a higher mass density driving the other.
This is especially true at large scales, where the effect of viscous dissipation vanishes %disappears
and the discrete nature of quantum vortices becomes indistinct. At such large inertial scales, there is consensus that the normal fluid and the superfluid are strongly tied and exhibit a joint {Kolmogorov's energy cascade}, regardless of temperature~\cite{Roche2009}. 
However, this locking cannot pertain down to the smallest scales. The two components must inevitably separate because they are not subject to the same viscous forces. Mutual friction then acts to reduce separation by exchanging momentum between the two fluids to compensate for this (viscous) imbalance \cite{Roche2009,Zhang2023}. 
Although this qualitative picture tends to be accepted, a quantitative understanding of how mutual friction affects small-scale dynamics, and possibly alters the definition of control parameters and characteristic scales, is still lacking \cite{Barenghi2014}.
Furthermore, controversy has recently arisen concerning
the possible effect of mutual friction on the intermittency phenomenon.
Intermittency here refers to the increasingly non-Gaussian form of velocity-difference distributions as the separation scale decreases~\cite{Frisch1995}.
Numerical investigations based 
on simulations of the {Hall-Vinen-Bekharevich-Khalatnikhov}~(\HVBK{}) {two-fluid model} have highlighted an enhancement \cite{Boue2013,Biferale2018,Verma2023} or a decrease \cite{Shukla2016} of intermittency
compared to the classical case 
at temperatures 
where the densities of normal fluid and superfluid are close ($\rho_n\approx \rho_s$) and thus the strength of mutual friction per unit mass (proportional to $\rho_n\rho_s/\rho^2$) is at a maximum. Intermittency enhancement aligns with numerical results obtained using
the Gross--Pitaevskii model~\cite{Krstulovic2016} (which does not account for finite temperature effects)
and with 
one experiment~\cite{Varga2018a}. It also aligns with a second experiment where the flow was driven by a heater~\cite{Bao2018}, and not by mechanical means as considered in the present study.
However, this enhancement is in total contradiction with experimental studies carried out at large Reynolds numbers~\cite{Maurer1998, Salort2011a, Rusaouen2017} and numerical studies using the \HVBK{} model~\cite{Salort2011a,Muller2022b}, which report intermittency properties invariant over a broad range of temperatures, and similar to those observed in classical turbulence. 

In what follows, the role of mutual friction is examined in the context of a coarse-grained
representation of the two-fluid He-II dynamics. Unlike previous studies~\cite{Verma2023}, the focus is on the behavior of the mixture rather than on individual fluid components. 
Our approach builds on and extends established tools from classical turbulence phenomenology -- such as dimensional analysis, scale-by-scale energy budget and characteristic scales -- to develop a robust framework for analyzing turbulence in He-II flows.
In particular, we derive an analogue of the classical {K\'arm\'an-Howarth equation} \cite{Frisch1995} suitable for the two-fluid mixture. 
This exact equation provides insights into the role of mutual friction in energy dissipation across scales, and motivates the introduction of an effective Reynolds number along with appropriate characteristic scales. 
Taking characteristic scales into account helps to resolve the apparent controversy regarding the temperature dependence of intermittency.

\section*{Coarse-grained two-fluid dynamics of He-II}
The most generally accepted equations for the coarse-grained dynamics of He-II are the so-called \HVBK{} equations \cite{barenghi_donnelly_2001}. 
The ``coarse-grained'' property relates to dynamics at scales significantly larger than the typical spacing ($\ell$) between quantum vortices. 
Therefore, the superfluid vorticity, which is singular by nature, is here approximated as a continuum, assuming a high density of underlying vortex lines at the resolved scales of motion. 
The \HVBK{} equations are 
\begin{align}
  \label{eq:HVBK}
  \begin{split}
    \diffp{\vn}{t} + (\vn \cdot \gradient) \vn &=
    - \frac{1}{\rhon} {\gradient \pn}
    + \frac{\rhos}{\rho} \Fns
    + \nun \laplacian \vn
    \\
    \diffp{\vs}{t} + (\vs \cdot \gradient) \vs &=
    - \frac{1}{\rhos} {\gradient \ps}
    - \frac{\rhon}{\rho} \Fns
    + \nus \laplacian \vs
  \end{split}
\end{align}
where $\pn$ and $\ps$ are the respective effective pressures 
and $\rho$ denotes the total density $\rhon + \rhos$. Isothermal dynamics
are associated with constant values of $\rhon$ and $\rhos$.
An ad-hoc kinematic viscosity $\nus$ is introduced for the superfluid to account for underlying dissipative phenomena arising from vortex reconnections {or vortex-line instabilities %(Kelvin-wave cascade)
} 
at scales smaller than the coarse-graining scale
\cite{Krstulovic2012}.  
In practice, $\nus$ is supposed (and fixed arbitrarily) much smaller than the viscosity $\nun$ of the normal fluid. 
Finally, the coupling between the two fluids per unit of volume occurs through the mutual friction force $(\rhon \rhos/\rho) \Fns$.
In what follows, the friction will be simplified to its main contribution as 
$\Fns \approx -\frac{B}{2} \kappa \VLD (\vn - \vs)$ with %$\VLD \approx \ell^{-2}$ 
$\VLD$
denoting the local length of vortex lines per unit volume, and with $B$ being an empirical coefficient of order unity that depends weakly on temperature~\cite{Barenghi1983}.
In the realm of the \HVBK{} model, $\kappa \VLD$ 
is approximated by the magnitude of the (coarse-grained) superfluid vorticity $\vortS$, giving
\begin{equation}
    \Fns = -\frac{B}{2} |\vortS| \, \vns
    \label{eq:Fns_hvbk}
\end{equation}
with  the slip velocity $\vns = \vn - \vs$ and $\vortS = \gradient \times \vs$.

Initially dedicated to the description of vortex waves in rotating He-II \cite{Henderson_2004} and the prediction of instabilities in Taylor-Couette flow \cite{barenghi_jones_1988,henderson_barenghi_2004}, the \HVBK{} equations also found applications in investigating He-II turbulence at finite temperatures~\cite{Roche2009, Boue2015, Shukla2015, Biferale2019a, Polanco2020a}.
An essential aspect of the \HVBK{} approach is to smooth the discrete and complex arrangement of quantum vortices,
and to represent the motion of the two fluids (at scales larger than $\ell$) in a self-consistent fluid-mechanical framework \cite{Barenghi2014}.
A known drawback is that 
the structure of the vortex tangle, its degree of local anisotropy, is not taken into account in the coarse-grained superfluid vorticity, which leads to some underestimation of the actual intensity of the friction force \cite{Nemirovskii2019a,Nemirovskii2020,Nemirovskii2021}.

The present work relies on the standard \HVBK{} model despite its limitations, as this model has facilitated the development of a substantial body of knowledge and interpretations regarding observations in superfluid $^4$He turbulence. The previously mentioned conflicting predictions about intermittency were derived from this model, which also highlights the importance of considering it. 
Finally, let us mention that, informed by previous studies \cite{Roche2009,Zhang2023}, a second-order $B^\prime$ term (Magnus term) and the vortex tension were ignored in the expression of the friction, thus retaining only the dominant contribution in the context of He-II turbulence.

\subsection*{Kármán-Howarth equation of He-II turbulence}

The K\'arm\'an-Howarth equation
is among the most important theoretical results in classical isotropic turbulence~\cite{Karman_Howarth_38,Frisch1995}.
{Derived from the Navier-Stokes equations, it provides, under the conditions of (local) homogeneity, isotropy and stationarity, an exact statistical signature of the balance between
  non-linear energy transfer and viscous dissipation at any scale of motion much smaller than the energy injection scale $L_f$.
When integrated over a ball of radius $r$,
the K\'arm\'an-Howarth equation yields the {scale-by-scale energy budget}~\cite{Antonia1997} 
\begin{equation}
  \frac{4}{3} \diss r = -S_3(r) + 2\nu \diff{}{r} S_2(r),
  \label{eq:KH_NS}
\end{equation}
where $\diss$ is the viscous energy dissipation rate per unit mass.
Here, $S_2(r) = \Emean{|\dvvec(\xvec,\rvec)|^2}$ and $S_3(r) = \Emean{|\dvvec(\xvec,\rvec)|^2 \dvpara(\xvec,\rvec)}$, with $\dvvec(\xvec,\rvec)=\vvec(\xvec+\rvec)-\vvec(\xvec)$, are respectively the second and third-order structure functions.
Additionally, $\dvpara(\xvec,\rvec) = \dvvec(\xvec,\rvec) \cdot \rvec / r$ denotes the longitudinal component of $\dvvec(\xvec, \rvec)$.
The term accounting for the 
external forcing at scale $r$ is here neglected for the sake of simplicity, which is consistent with the hypothesis that the Reynolds number is sufficiently high and, therefore, that finite Reynolds number effects are negligible at such scale \cite{Zhang2023}.   
Importantly, at statistical equilibrium, the one-point energy budget reduces to $\diss_{\text{inj}} = \diss$,
where $\diss_{\text{inj}}$  denotes the rate of energy injection (per unit mass) at large scales, reflecting that the energy supplied to the fluid is ultimately dissipated through viscosity.
Therefore the left-hand-side of \eqref{eq:KH_NS}  is also equal to
$  \frac{4}{3} \diss_\mathrm{inj}  r$. 
In fact, the same quantity $\diss$ amounts to the energy injected at large scales, the flux of energy through the {inertial} range of scales and the energy dissipated at small scales. 

The scale-by-scale energy budget for He-II can be derived similarly, by combining the energy budgets obtained for the two fluids individually, based on \eqref{eq:HVBK}. 
Local homogeneity, isotropy, and stationarity are again assumed. 
This yields
\begin{equation}
  \label{eq:four_thirds_law}
  \frac{4}{3} \epsinj r = -\SSStilde(r) + 2\nutilde \diff{}{r} \SStilde(r) + \phiNS(r)
\end{equation}
where $\epsinj = (\rhon \epsinjn + \rhos \epsinjs)/\rho$ is the total energy injection rate ($\epsinjn$ and $\epsinjs$ being the injection rate for the two fluid components respectively) and $\nutilde = (\rhon \nun + \rhos \nus) / \rho$ can be viewed as an {effective} kinematic viscosity for the mixture.
Moreover, the structure functions in \eqref{eq:four_thirds_law} appear as 
mass and viscosity-weighted combinations of the individual structure functions of the two fluids, namely
$\Stilde_3(r) = (\rhon S_3^{\text{n}}(r) + \rhos S_3^{\text{s}}(r))/\rho$ and
$\Stilde_2(r) = (\rhon \nun S_2^{\text{n}}(r) + \rhos \nus S_2^{\text{s}}(r))/\rho \nutilde$.
  
The contribution of the mutual friction is encompassed in $\phiNS(r)$ with
\begin{equation}
\phiNS(r) = \frac{\rhon \rhos}{\rho^2} \, 
\frac{2}{r^{2}} \int_{0}^{r} \Cns(r') \, r'^2 \, \dd r'
\label{eq:phins_general}
\end{equation}
where $\Cns(r) \equiv \langle \vns(\xvec + \rvec)\cdot \Fns(\xvec) + \vns(\xvec) \cdot \Fns(\xvec+\rvec) \rangle$
is the two-point correlation between the slip velocity $\vns$ and the mutual friction force $\Fns$.
Finally, note that the left side of \eqref{eq:four_thirds_law} contains $\epsinj$, which differs from the viscous energy dissipation rate, unlike in classical turbulence at statistical equilibrium.
 
At this stage, Eq.~(\ref{eq:phins_general}) is very general and independent of the expression of the mutual friction. If \eqref{eq:Fns_hvbk} is now prescribed, one eventually obtains 
\begin{equation}
\phiNS(r) \approx A \, \kappa \Emean{\VLD} \, \frac{4}{r^{2}} \int_{0}^{r} \langle \vns(\xvec + \rvec')\cdot \vns(\xvec)\rangle  \, r'^2 \, \dd r'
\label{eq:phins}
\end{equation}
by replacing the vortex-line density with its mean value 
$\Emean{\VLD}$ in the integral,
i.e.~neglecting the fluctuations of $\VLD = |\vortS|/\kappa $ at the scale of variations of the slip velocity \cite{Roche2009}
and posing $A = (\rhon\rhos/\rho^2)B/2$. 
In the limit $r \to 0$, \eqref{eq:four_thirds_law} simplifies to the one-point energy budget
$\epsinj = \epsvisc + \epsMF$ with $\epsvisc = (\rhon \epsN + \rhos \epsS) / \rho$ 
and $\epsMF = A 
\kappa \Emean{\VLD}
\Emean{|\vns|^2}$ representing the viscous and frictional dissipation rate (per unit mass), respectively (see Materials and Methods).
This budget reflects that the energy supplied to the fluid is now dissipated from viscosity and internal friction between the two fluid components.

It is important to emphasize that \eqref{eq:four_thirds_law} holds for He-II {as a mixture} rather than for individual components.
This new perspective highlights the resemblance between the energy budget of classical turbulence (\eqref{eq:KH_NS}) and that of He-II turbulence (\eqref{eq:four_thirds_law}), 
but also the conditions under which they can differ, when $\phiNS(r)$ becomes predominant.
From its definition, this term is expected to be largest when $\rhon \approx \rhos$ and $A \propto \rhon \rhos /\rho^2$ reaches its peak value. Interestingly, this corresponds approximately to temperatures 
where previous controversies about intermittency arose. 

\section*{Characteristic microscales of He-II turbulence}

In three-dimensional He-II turbulence, each single-fluid component experiences both {an} energy cascade from large to small scales and an energy exchange with the other component caused by mutual friction \cite{Roche2009,Zhang2023}.
Consequently, the energy injected into a component is not necessarily dissipated at small scales by this same component, which leads us to reconsider the definitions of the Kolmogorov's and Taylor's microscales, and then of the Reynolds number.

\subsection*{Energy fluxes}

To illustrate the energetics of He-II turbulence, the different energy fluxes in scale space are first considered at different temperatures, for each component individually and for the two-fluid mixture.
The energy flux associated with the energy cascade (triadic interactions) is defined as $\Pi(k) = \Emean{\vvec^{<k} \cdot \left( \vvec \cdot \gradient \vvec \right)}$, where $\vvec^{<k}$ is the low-pass filtered velocity at wavenumber $k$~\cite{Frisch1995}.
In classical turbulence, and at scales smaller than the energy injection ones, this flux is compensated by the cumulative viscous dissipation $\mathcal{D}(k) = \nu \Emean{|\gradient \vvec^{<k}|^2}$ up to wavenumber $k$, so that 
\begin{equation}
\diss_{\text{inj}} = \Pi(k) + \mathcal{D}(k)
\end{equation}
which can be viewed as the analogue of \eqref{eq:KH_NS} in the spectral domain.
In the presence of mutual friction, an additional term $\mathcal{T}^{\sigma}(k)$ must be included in the energy balance associated with each component $\sigma \in \{\text{n}, \text{s}\}$ to account for the net energy exchange with the other component.
Namely, $\FluxTn(k) = (\rhos / \rho) \Emean{\vn^{<k} \cdot \Fns}$ and
$\FluxTs(k) = -(\rhon / \rho)\Emean{\vs^{<k} \cdot \Fns}$.
Therefore, the revised budget writes
$\epsinj^{\sigma} = \Pi^{\sigma}(k) + \mathcal{D}^{\sigma}(k) + \mathcal{T}^{\sigma}(k)$.
Finally, one obtains for the He-II mixture
\begin{equation}
    \epsinj = \widetilde{\Pi}(k) + \widetilde{\mathcal{D}}(k) + {\mathcal{T}}_\text{ns}(k)
\end{equation}
where the tildes indicate mass-weighted averages over the two fluids. The overall exchange of energy 
${\mathcal{T}}_\text{ns}(k) = (\rhon/\rho) \mathcal{T}^n(k) + 
(\rhos/\rho) \mathcal{T}^s(k)$ results in the dissipation by mutual friction which can be related to $\phiNS(r)$ in \eqref{eq:four_thirds_law}.

The above fluxes are evaluated using direct numerical simulations of the HVBK equations (see Materials and Methods).
For individual components, as illustrated in the left and middle columns of Fig.~\ref{fig:energy_fluxes}, friction-induced energy exchanges (green dotted lines) are prominent at small scales. The amplitudes of these exchanges strongly depend on temperature and are comparable in absolute value to the viscous dissipation (orange dashed lines).
Friction acts as an energy sink for the superfluid at all considered temperatures. This effect intensifies as the temperature approaches $T_\lambda$.
In contrast, friction operates as an energy source for the normal fluid, and this contribution increases as the temperature decreases from $T_\lambda$. 
This observation is consistent with the idea that friction tends to bind the two fluids by compensating for their differences in kinematic viscosity, and that the mixture 
appears to be driven by the fluid component with the highest mass ratio \cite{Zhang2023}.
However, when considering the mixture (right column of Fig.\ref{fig:energy_fluxes}), the remarkable observation is that, regardless of the temperature, the overall kinetic effect of mutual friction remains a sub-dominant contribution to the viscous dissipation. Consequently, the emerging picture resembles classical turbulence with a
slightly enhanced dissipation at very small scales due to friction. This later calls for revising the characteristic microscales to account for this increased dissipation. 

\begin{figure*}[!t]
  \centering
  \includegraphics[width=\linewidth]{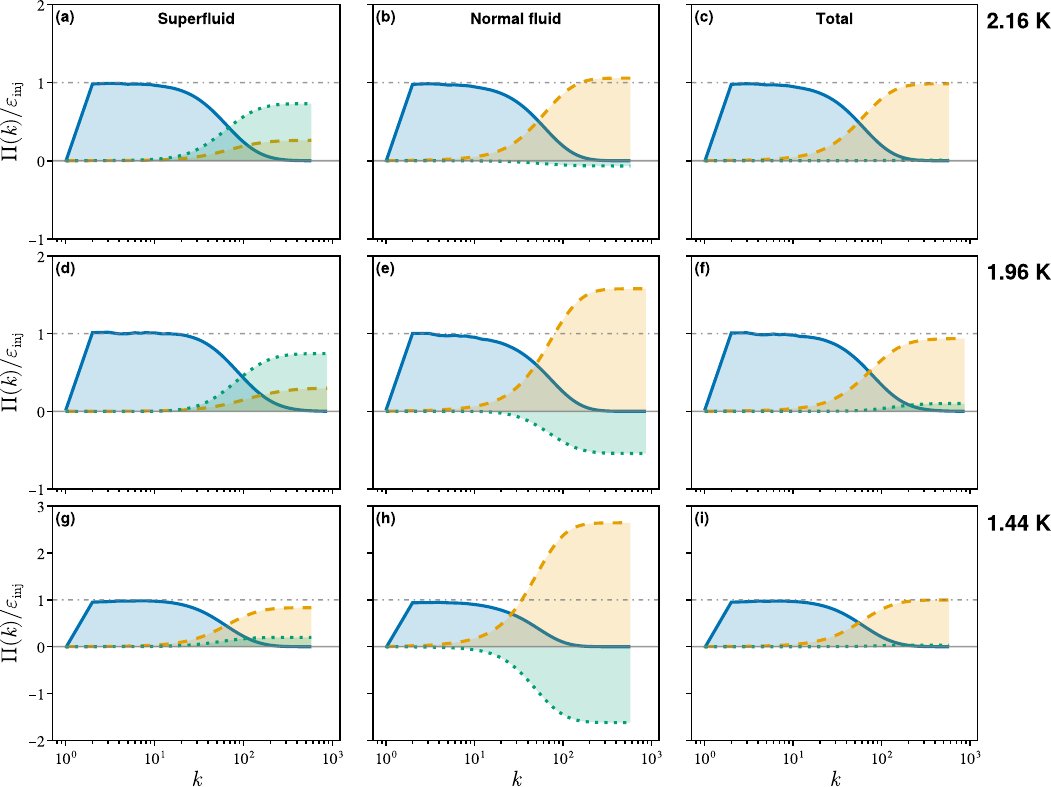}
  \caption{%
    Normalized energy fluxes across wavenumbers $k$ for the superfluid (left column), normal fluid (middle column) and two-fluid mixture (right column).
    Blue solid lines, energy-cascade flux $\Pi(k)$;
    orange dashed lines, cumulative viscous dissipation rate $\mathcal{D}(k)$;
    green dotted lines, inter-component energy flux $\mathcal{T}^\sigma(k)$ due to mutual friction (left and middle columns) and related cumulative dissipation rate ${\mathcal{T}}_\text{ns}(k)$ (right column).
    Obtained from \HVBK{} simulations at $T\approx 2.16\,\text{K}$ (top), $T \approx  1.96\,\text{K}$ (middle) and $T\approx 1.44\,\text{K}$ (bottom) corresponding respectively to \CaseThighAdapted{}, \CaseTmidExtTer{} and  \CaseTlowTer{} in Table~\ref{tab:DNS_parameters}.
  }
  \label{fig:energy_fluxes}
\end{figure*}

\begin{table*}[tb]
  \centering
  \caption{%
    Parameters and characteristic scales of simulations at various temperatures.
    The first column indicates the symbol representing each simulation in the figures.
    $N$ is the grid size in each direction.
    The last row is a simulation of the incompressible Navier-Stokes equations.
    See main text for other definitions.
  }
  \label{tab:DNS_parameters}
  % \begin{ruledtabular}
  \setlength{\tabcolsep}{5mm}
  \begin{tabular}{ccccccccccc}
    \toprule
          & $T$ (K) & $\rhos / \rhon$ & $\nus / \nun$ & $B$ & $N$
          & $\kmax \etatilde$ & $\ell / \etatilde$ & $\lambdatilde / \etatilde$ & $L_0 / \lambdatilde$ & $R_{\tilde\lambda}$
    \\
    \midrule
    \CaseTlowBis{}      & 1.44  & 10  & 0.2          & 2.0  & 1152 & 1.6 & 2.09 & 33.5 & 19.4 & 294 \\
    \CaseTlowTer{}      & 1.44  & 10  & 0.2          & 2.0  & 1152 & 2.9 & 2.07 & 26.8 & 12.5 & 192 \\
    \CaseTmidExt{}      & 1.96  & 1   & 0.2          & 1.0  & 1728 & 3.8 & 3.04 & 28.4 & 14.0 & 225 \\
    \CaseTmidExtBis{}   & 1.96  & 1   & 0.2          & 1.0  & 1728 & 3.1 & 3.04 & 30.6 & 16.2 & 260 \\
    \CaseTmidExtTer{}   & 1.96  & 1   & \textbf{0.1} & 1.0  & 1728 & 3.6 & 3.06 & 29.0 & 14.5 & 232 \\
    \CaseThighAdapted{} & 2.157 & 0.1 & 0.2          & 2.16 & 1152 & 2.9 & 2.80 & 27.6 & 13.1 & 203 \\
    \CaseNS{}           & NS    & 0   & --           & --   & 2048 & 1.4 & --   & 44.6 & 34.2 & 510 \\
    \bottomrule
  \end{tabular}
  % \end{ruledtabular}
\end{table*}

\subsection*{Taylor's and Kolmogorov's microscales}

In isotropic classical turbulence, the Taylor microscale $\lambda$ is defined from the dissipation rate $\diss$ via the relation~\cite{Tennekes72}
\begin{equation}
	\diss = {15 \nu} \frac{v_\text{rms}^2}{\lambda^2} 
	\label{eq:def_lambda}
\end{equation}
where $v_\mathrm{rms}^2 = \Emean{|\vvec|^2} / 3$.
This equality makes use of the exact expression $\diss = 15 \nu \left < (\partial v_x/\partial x)^2 \right >$, which holds in isotropic turbulence. 
The Taylor's microscale then defines the Reynolds number $\Rlambda = \vrms \lambda / \nu$, which is a key parameter used to quantify finite Reynolds number effects in turbulence, see e.g.~\cite{Antonia2017, Sinhuber2017, Elsinga2020, Iyer2020a, Buaria2022b}.

%The definition of $\lambda$ for two-fluid quantum turbulence is less clear. 
Defining $\lambda$ for two-fluid quantum turbulence is less straightforward.
One
possible approach involves establishing separate $\lambda$'s (as well as the corresponding $\Rlambda$'s) for each fluid component~\cite{Biferale2018, Muller2022b}. However, this lacks a clear physical interpretation due to inter-component energy exchanges by mutual friction. Therefore, it may not offer a meaningful insight when comparing \HeFour{} turbulence
at different temperatures. 
A natural alternative, based on our previous two-fluid analysis, is to 
use $\epsvisc$ (in replacement of $\diss$ in classical turbulence) as an intermediary to define a Taylor's microscale for the mixture.
This leads to a generalization of \eqref{eq:def_lambda} applicable to two-fluid dynamics,
\begin{equation*}
\epsvisc = 15 \, \nutilde \,\frac{\vrms^2}{{\widetilde{\lambda}}^2}
\end{equation*}
in which $3 \vrms^2 = \Emean{|\vn|^2} \approx \Emean{|\vs|^2}$ due to the strong locking of the two fluids at large scales.
Equivalently, one obtains that
\begin{equation}    
\frac{1}{\widetilde \lambda^2} = \frac{{\rhon \nu_n}/{\lambda_n^2}
+ {\rhos \nu_s }/{\lambda_s^2}}{\rho_n \nu_n + \rho_s \nu_s},
\label{eq:def_lambda_mixture}
\end{equation}
where the Taylor's microscales $\lambda_\sigma$ associated to each fluid ($\sigma = \text{n,\,s}$) are defined based on their respective viscous dissipations $\diss_\sigma$.
Therefore, \eqref{eq:def_lambda_mixture} provides a unique Taylor's microscale for the mixture dynamics, based on the exact energy budget~\eqref{eq:four_thirds_law}.
Moreover, it allows us to consistently define the Taylor-based Reynolds number for the two-fluid mixture as $\RlambdaTilde = \vrms \lambdatilde / \nutilde$.
We emphasize that in \eqref{eq:def_lambda_mixture}, we use the viscous dissipation $\epsvisc$ instead of the total dissipation $\epsvisc + \epsMF$ since only the viscous dissipation is a proxy to the variance of the velocity gradients.

In classical turbulence, Kolmogorov's microscale $\eta$ characterizes the cross-over between the inertial and dissipative dynamics. It can be estimated by equalling the expressions of the second-order structure function in the inertial and dissipative ranges. 
Note that this assumes that the turbulence is sufficiently developed to exhibit a separation between the inertial and dissipation scales.
In the context of two-fluid quantum turbulence, the same approach yields
\begin{equation}
 {\epsinj}^{2/3}r^{2/3} \approx \frac{{\epsvisc}}{\nutilde} \, r^2 \quad \textrm{for}~r = \widetilde{\eta},
\end{equation}
which leads to 
\begin{equation}
\widetilde{\eta} = \left( \frac{\nutot^3}{\epsinj} \right)^{1/4} 
\label{eq:kolmogorov_scale}
\end{equation}
by considering
\begin{equation} \nutot = \nutilde \, \frac{\epsinj}{\epsvisc}. \label{eq:overall_viscosity}\end{equation} 
This estimate echoes the expression appearing in classical turbulence with the particularity that $\nutot$ should be interpreted here as an effective viscosity for the mixture.
Pushing this phenomenological approach further, we next derive an explicit prediction for the characteristic distance between quantized vortices, thereby bridging quantum and hydrodynamical features of He-II turbulence.

\subsection*{Inter-vortex distance}

\begin{figure}
    \centering
    \includegraphics[width = 0.9\linewidth]{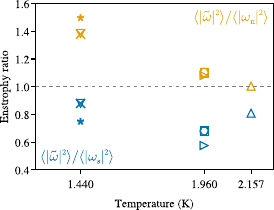}
    \caption{%
    Ratio between the effective vorticity variance $\Emean{|\tilde{\vortvec}|^2}$ and the vorticity variance $\Emean{|\vortvec_{\sigma}|^2}$ related to  each component of He-II  ($\sigma = \text{n}, \text{s}$).
    The simulations listed in Table~\ref{tab:DNS_parameters} are represented with their corresponding symbols.
    Asterisks correspond to an additional simulation with $\nus/\nun = 0.1$ at $T \approx 1.44$\,K.
    }
    \label{fig:enstrophy_ratio}
\end{figure}

Following the introduction of an effective viscosity for the mixture, \eqref{eq:overall_viscosity}, prompts us to \emph{formally} introduce a 
``viscosity'' associated with mutual friction as 
\begin{equation}
    \nuMF \equiv \nutot - \nutilde = \nutilde \,\frac{\epsMF}{\epsvisc}, 
    \label{eq:def_nufriction}
\end{equation}
so that 
$\epsMF  =\nuMF \left( {\epsvisc}/{\nutilde}\right) =\nuMF \, \langle |\widetilde{\vortvec}|^2 \rangle$
with
\begin{equation}
    \langle |\widetilde{\vortvec}|^2 \rangle = \frac{\epsvisc}{\nutilde} = \frac{\rhon \nun \langle |\vortN|^2 \rangle+ \rhos \nus \langle |\vortS|^2 \rangle}{\rhon \nun + \rhos \nus}.
\end{equation}
The ratios $\langle |\widetilde{\vortvec}|^2 \rangle / \langle |\vortN|^2 \rangle$ and $\langle |\vortvec|^2 \rangle / \langle |\vortS|^2 \rangle$ are displayed in  Fig.~\ref{fig:enstrophy_ratio}. %for the different simulations. 
We see that the ratio $C_\omega \equiv \langle |\widetilde{\vortvec}|^2 \rangle / \langle |{\vortS}|^2 \rangle$ is a unity order factor which depends weakly on the temperature and viscosity ratio $\nus/\nun$, and remains independent of the Reynolds number.
Therefore, by plugging $\epsMF = \nuMF\, C_\omega\, \langle |{\vortS}|^2 \rangle$ in \eqref{eq:def_nufriction} and incorporating the \HVBK{} assumption $|\vortS| = \kappa \VLD$, one gets that
\begin{equation}
    \epsvisc = C_\omega\,\nutilde \kappa^2 \langle \VLD^2 \rangle
\end{equation}
which can be rewritten as $\epsvisc = 2\,C_\omega\,\nutilde \kappa^2 /\ell^4$ by introducing the characteristic inter-vortex-line spacing $\ell = 1 / \langle \VLD \rangle^{1/2}$ and  taking $\langle \VLD^2 \rangle = 2 \langle \VLD \rangle^2$ as argued in~\cite{Roche2008}. 
If we further assume that 
the dissipation by mutual friction remains small compared to the viscous dissipation (see Fig.~\ref{fig:energy_fluxes}),
one obtains that  $ \epsinj \approx \epsvisc = 2\,C_\omega\,\nutilde \kappa^2/ \ell^4$.
Therefore, writing $\epsinj = {C_\epsilon} \vrms^3/L_0$ where $L_0$ represents the integral scale {and $C_\epsilon$ is the dissipation constant}, leads to the vortex spacing $\ell_\mathrm{hvbk}$ in the context of the \HVBK{} modeling
\begin{equation}
    \frac{\ell_\mathrm{hvbk}}{L_0} \approx 2^{1/4} \left(\frac{C_\omega}{{C_\epsilon}} \, \frac{\frac{\mu}{\rho}}{\kappa} \right)^{1/4} \, \mathrm{Re}_\kappa^{-3/4} 
    \label{eq:intervortex_law}
\end{equation}
with $\mathrm{Re}_\kappa = \vrms L_0/\kappa$ and 
$\mu/\rho = \nutilde \approx 9.7(\pm0.8) \, 10^{-9} \, \mathrm{m^2/s}$ for the kinematic viscosity of He-II in the range $1.3\, \mathrm{K} \leq T \leq 2\,\mathrm{K}$~\cite{Donnelly1998}.

Further refinement is possible
by taking into account the fact 
that \HVBK{} modeling underestimates the vortex-line density by omitting the contribution of unpolarized vortices, leading to an overestimation of $\ell$ \cite{Roche2007, Roche2009}. The ratio $(\ell/\ell_\mathrm{hvbk})^2$ has been estimated to be about $1/2$ in isotropic He-II turbulence \cite{Roche2009}, which finally leads to  
\begin{equation}
\frac{\ell}{L_0} \approx \frac{1}{2^{1/4}} \left(\frac{C_\omega}{{C_\epsilon}} \,\frac{\frac{\mu}{\rho}}{\kappa} \right)^{1/4} \! \mathrm{Re}_\kappa^{-3/4}
\label{eq:model}
\end{equation}
and eventually
\begin{equation}
\frac{\ell}{L_0}
\approx {0.5} \, \mathrm{Re}_\kappa^{-3/4}
\label{eq:coef05}
\end{equation}
by considering
$C_\omega^{1/4} 
\approx 1$ within the range of temperatures $1.44~\mathrm{K} \lesssim T  \lesssim 2~\mathrm{K}$ (see Fig.~\ref{fig:enstrophy_ratio}) and
$C_\epsilon^{1/4} \approx 1$
in the context of classical isotropic turbulence \cite{vassilicos}. 

\begin{figure}
    \centering
    \includegraphics[width = 1.0\linewidth]{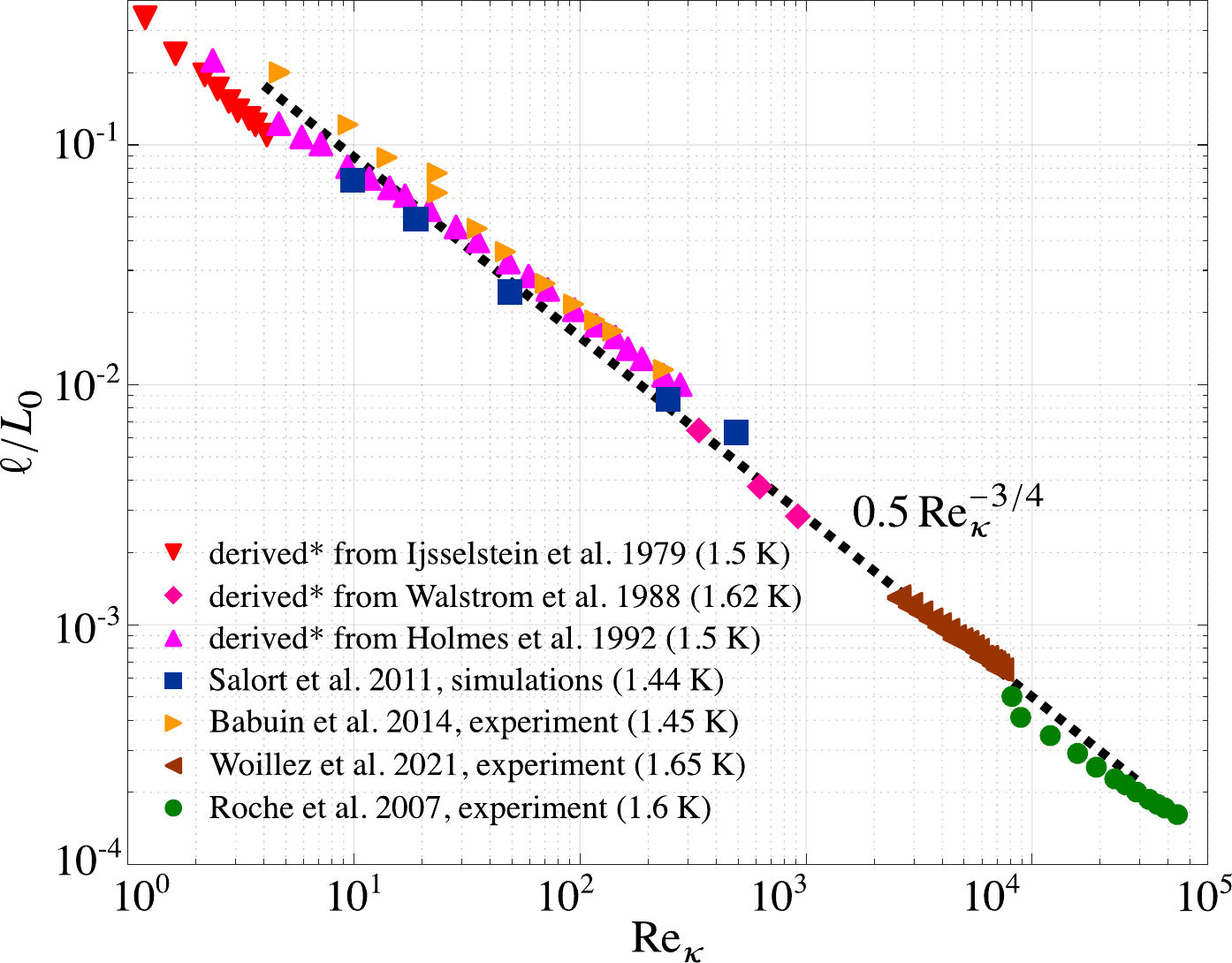}
    \caption{%
    Inter-vortex distance normalized by the estimated flow integral scale in various steady flow experiments and simulations around $1.55 \pm 0.1$\,K. Data from Ijsselstein \emph{et al.} \cite{Ijsselstein1979}, Walstrom \emph{et al.} \cite{Walstrom1988b} and Holmes \emph{et al.} \cite{Holmes1992} have been processed under the same assumptions as Salort \emph{et al.} \cite{Salort2011}. Namely, these pipe flows produce turbulence with an intensity of 5\% and an integral scale twice the transverse flow scale (four times for Holmes \emph{et al.}). In Roche \emph{et al.} \cite{Roche2007}, both quantities were derived from velocity spectrum measurements. Finally, measurements in Babuin \emph{et al.} 
 \cite{Babuin2014} and Woillez \emph{et al.} \cite{Woillez2021} were performed in the near wake of grids, in conditions where a turbulent intensity of about $9\%$ was expected \cite{Babuin2014,Woillez2021}. The integral scale was respectively estimated from the literature \cite{Babuin2014} and from velocity spectra \cite{Woillez2021}.
    }
    \label{fig:InterVortex_vs_Re}
\end{figure}

\eqref{eq:coef05} has a long history but, to the best of our knowledge, had never been derived from the \HVBK{} equations. It was first proposed in 2006~\cite{RocheFlorida2006} by analogy with classical turbulence, where it was argued that the Kolmogorov viscous scale $\eta$ normalized by the integral length scale $L_0$ follows a similar relation with the Reynolds number, i.e.~$\eta/L_0\propto \mathrm{Re}^{-3/4}$. 
This proposal permitted a reinterpretation of previously published data~\cite{Ijsselstein1979,Walstrom1988b,Holmes1992} and other data reported the following year~\cite{Roche2007}. The validity of \eqref{eq:coef05} around $1.5 - 1.6$\,K was confirmed in 2011 by numerical simulations using the \HVBK{} model~\cite{Salort2011}. In 2014~\cite{Babuin2014}, an equivalence was shown between the empirical equation~\eqref{eq:coef05} and the empirical proportionality relation $\diss = \nu_{\text{eff}} \left( \kappa \langle \VLD \rangle \right) ^2$, which was postulated in~\cite{Smith1993,Vinen2002} and widely used to interpret the experiments of He-II turbulence decay (see \cite{Smith1993,stalp2002,Niemela:JLTP2005,WalmsleyPNAS2014,babuin2014decay} for instance). The data cited above, and some others discussed later, are presented in Figs.~\ref{fig:InterVortex_vs_Re} and \ref{fig:InterVortex_vs_T}. 
The predicted scaling law $\ell/L_0 \approx 0.5\,\mathrm{Re}_\kappa^{-3/4}$ is found in excellent agreement with all experimental measurements over more than four decades of Reynolds numbers.
Notably, this scaling extends down to $\mathrm{Re}_\kappa$ values of order unity.

\begin{figure}
    \centering
    \includegraphics[width = 1.0\linewidth]{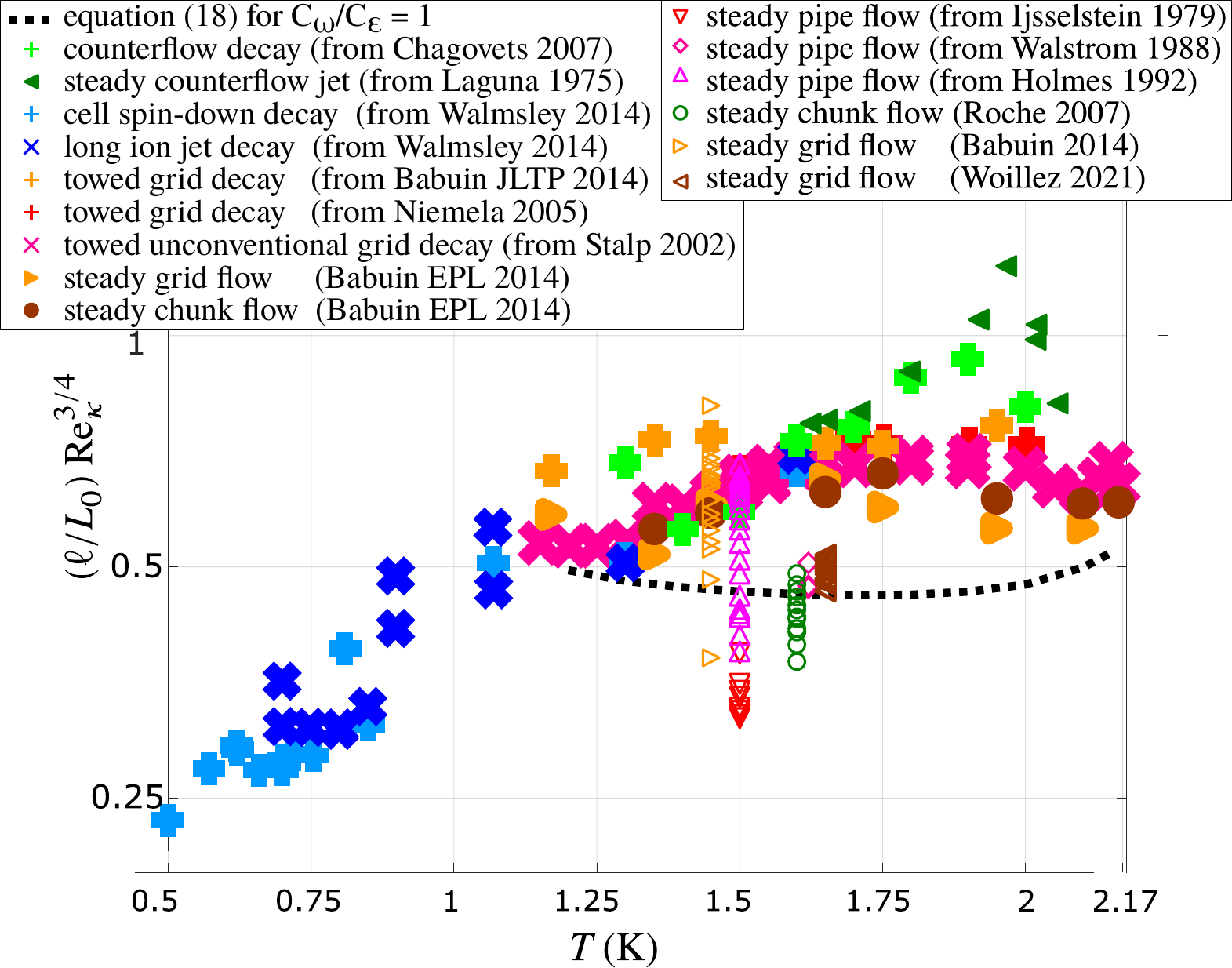}
    \caption{%
    Normalized inter-vortex distance relative to the estimated flow integral scale and compensated by the Reynolds number in various experiments. The open symbols (right-side legend) correspond to the steady flow datasets shown in Fig.~\ref{fig:InterVortex_vs_Re}.
    Additional steady flows are a grid flow~\cite{Babuin2014} (orange triangles), an ill-defined pipe flow~\cite{Babuin2014} (brown discs), and a counterflow jet~\cite{LagunaPRB1975} (dark green triangles). The effective viscosity $\nu_{\textup{eff}}$ derived from {turbulent He-II decay experiments} allows us to derive an inter-vortex distance following \eqref{eq:coefNuEff} and assuming $C_\epsilon=1$. For such experiments, the compensated normalized inter-vortex distance reported along the vertical axis is simply $\left( {\nu_{\text{eff}}}/{\kappa C_\epsilon} \right )^{1/4}$. Plus and cross symbols correspond to such decaying experiments and were obtained in an initially towed-grid flow~\cite{Niemela:JLTP2005,stalp2002} (red and pink), spinning down cell~\cite{WalmsleyPNAS2014} (blue aqua), a flow initially forced by an ion jet~\cite{WalmsleyPNAS2014} (blue) and in a decaying counterflow~\cite{Chagovets:PRE2007} (chartreuse green).
    }
    \label{fig:InterVortex_vs_T}
\end{figure}

The value of the prefactor in \eqref{eq:coef05} and its temperature dependence within the range of interest are examined in Fig.~\ref{fig:InterVortex_vs_T}. 
The open symbols correspond to the data displayed in Fig.~\ref{fig:InterVortex_vs_Re} within the temperature range of 1.45–1.65\,K. 
Across all considered experiments, the prefactor remains within $0.6\pm 0.3$ over more than four decades in $\mathrm{Re}_\kappa$. 
Part of the dispersion may be attributed to uncertainties in the estimation of the integral scale and the turbulence level, as the prefactor depends on $L_0^{-1/4}$ and $\vrms^{3/4}$. Another possible source of errors 
is
the lack of isotropy and homogeneity of the turbulence produced in some less well-defined flows~\cite{Ijsselstein1979,Walstrom1988b,Holmes1992,Roche2007}.
The flow dependence of the dissipation ``constant'' $C_\epsilon$, which can reach values as low as 0.5 in the near wake of a grid~\cite{vassilicos}, can result in a 19\% increase of the prefactor in \eqref{eq:model}.
Interestingly, the steady grid flow reported in~\cite{Woillez2021} provides a dataset for which $L_0$ and $\vrms$ have been accurately measured in situ in a well-defined flow at high $\mathrm{Re}_\kappa$. The prefactor estimate is about $0.5$, fully confirming \eqref{eq:coef05}.
The temperature dependence is revealed more clearly by considering the same flow at different temperatures within the range of interest. The dataset labeled ``steady grid flow (Babuin EPL 2014)''~\cite{Babuin2014} fulfills this requirement and shows little or no dependence on temperature between 1.17\,K and 2.16\,K. Although the value of the prefactor relies on estimates of $L_0$ and $\vrms$ from the literature, these estimates should not vary with temperature at a fixed position in the wake of the grid.
For completeness, the plus and cross symbols in Fig.~\ref{fig:InterVortex_vs_T} correspond to datasets obtained from decaying turbulent flows, rather than stationary flows as considered so far.

\subsection*{Effective viscosity of He-II}

Decaying He-II turbulent flows are usually analyzed under the assumption that the total dissipation rate, or equivalently $\epsinj$, follows the relation~\cite{Smith1993,Vinen2002}
\begin{equation}
    \epsinj = \nu_{\text{eff}} \left( \kappa \langle \VLD \rangle \right) ^2,
    \label{eq:def_nueff}
\end{equation} and that the integral scale $L_0$ remains constant during the late decay. 
This  assumption allows us to determine an effective viscosity $\nu_{\text{eff}}$ for He-II turbulent flows \cite{Smith1993}.
By considering $\epsinj =  {C_\epsilon} \vrms^3/L_0$, the definition of $\nu_{\text{eff}}$ implies
\begin{equation}
\frac{\ell}{L_0}
= \left( \frac{\nu_{\text{eff}}}{\kappa {C_\epsilon}}  \right) ^{1/4} \mathrm{Re}_\kappa^{-3/4}.
\label{eq:coefNuEff}
\end{equation}
This reformulation (with ${C_\epsilon}\approx 1$) has allowed us to include datasets from decay experiments in Fig.~\ref{fig:InterVortex_vs_T}. These datasets support the weak temperature dependence of the prefactor observed in steady flows above approximately 1.17\,K, while also demonstrating its decrease at lower temperatures \cite{WalmsleyPNAS2014}. This decrease, which falls beyond the range of applicability of the \HVBK{} model and therefore, of the present study, will be discussed below.
Let us mention that, in practice, $\nu_{\text{eff}}$ is only determined up to a prefactor of order one, mainly due to its proportionality to $L_0^2$, which is most often assumed to correspond exactly to the size of the flow vessel during the late decay phase. This assumption could lead to a systematic overestimation of the effective viscosity $\nu_{\text{eff}}$ deduced from decay experiments.
Incidentally, an alternative way to formulate \eqref{eq:model} is to provide a prediction for the effective viscosity $\nu_\textup{eff}$ by identification of \eqref{eq:model} with \eqref{eq:coefNuEff}, namely
\begin{equation}
 \nu_{\textup{eff}} = \frac{C_\omega}{2} \,\frac{\mu}{\rho}.
 \label{eq:modelNuEff}
\end{equation}
Emphasis is now placed on the low-temperature regime, particularly at temperatures $T \lesssim 1~\mathrm{K}$ for which $\rho_n/\rho \lesssim 1\% $, and on the upper limit as $T$ approaches $T_\lambda$.
In these regimes, the concentration of one fluid component largely dominates the other, making the \HVBK{} description less relevant.

As illustrated by the blue crosses and pluses~\cite{WalmsleyPNAS2014} in Fig.~\ref{fig:InterVortex_vs_T}, the inter-vortex spacing $\ell / L_0$ decreases as the temperature is reduced 
down to $0.5\,\mathrm{K}$. 
This pronounced temperature dependence can be explained by assuming that dissipation is predominantly due to mutual friction. 
Following the model $\nu_{\textup{eff}} = \frac{\rho_s \rho_n}{\rho^2} \, \frac{B}{2}\kappa$ presented in~\cite{Babuin2014}, 
one obtains
\begin{equation}
\frac{\ell}{L_0}
= \left(\frac{\rho_s \rho_n}{\rho^2} \, \frac{B}{2 {C_\epsilon}}  \right) ^{1/4} ~\mathrm{Re}_\kappa^{-3/4}\quad \mathrm{for}\,\, T \lesssim 1.2~\mathrm{K}.
\end{equation}
Here, the strong temperature dependence at low temperatures comes from the factor $\rho_s \rho_n / \rho^2$. 
The agreement of this model (within a numerical factor of order unity) with low-temperature data, therefore, suggests a crossover in the dominant dissipation mechanism in turbulent He-II, from viscous dissipation at intermediate temperatures, as shown by the present study, to the dissipation by mutual friction below approximately $1.2\,\mathrm{K}$ as supported by~\cite{Babuin2014}.

In the high-temperature limit, the normal fluid dynamics as well as the energy dissipation in He-II are not affected by the negligible superfluid concentration, so we expect $\epsinj = \frac{\mu}{\rho} \langle |\vortN|^2 \rangle$.
By using \eqref{eq:def_nueff}
and $\kappa \langle \VLD \rangle =
(\ell_\mathrm{hvbk} / \ell)^2  \langle |\vortS| \rangle$, one gets
\begin{equation}
\nu_{\text{eff}} = \frac{\mu}{\rho}\,  \frac{\langle |\vortN|^2 \rangle}{ \langle |\vortS| \rangle^2 } \left( \frac{ \ell}{\ell_\mathrm{hvbk}} \right)^4 \quad  \mathrm{as}\,\, T \rightarrow T_\lambda.
\end{equation}
The estimates $\langle |\vortN|^2 \rangle \approx 2 \langle |\vortN| \rangle ^2$~\cite{Roche2008} and
$(\ell/\ell_\mathrm{hvbk})^2 \approx 1/2$~\cite{Roche2009}, and further assuming that the locking of the dilute superfluid to the normal fluid in the high-temperature limit implies that $\langle |\vortN| \rangle \approx \langle |\vortS| \rangle$, eventually give 
\begin{equation}
  \nu_{\text{eff}} \approx \frac{\mu}{2 \rho} \quad \mathrm{as}\,\, T \rightarrow T_\lambda, \label{eq:nu_eff_high_T}
\end{equation}
which is consistent with our \eqref{eq:modelNuEff} by taking $C_\omega =1$ in this limit (see Fig.~\ref{fig:enstrophy_ratio}).

One might find surprising the one-half factor in \eqref{eq:nu_eff_high_T}.
However, this is a consequence of the original definition $\nu_{\text{eff}} = \epsinj / (\kappa \langle \VLD \rangle)^{2}$. 
This definition was proposed by analogy with the relation $\nu = \epsilon / \langle |\vortvec|^2 \rangle$ in classical turbulence, but the analogy would have been more relevant if
we had considered $\langle (\kappa \VLD)^2 \rangle$ as a surrogate for superfluid enstrophy, rather than $\langle \kappa \VLD \rangle^2$.
Consequently, the effective viscosity $\nu_{\text{eff}} $ defined in He-II does not correspond to the kinematic viscosity $\mu/\rho$ of He-I when the superfluid density vanishes. To our knowledge, this point had never been raised before in the literature.

\subsection*{Additional remarks}

Fig.~\ref{fig:InterVortex_vs_T} includes two independent datasets, labeled ``counterflow decay (from Chagovets 2007)''~\cite{Chagovets:PRE2007} and 
``steady counterflow jet (from Laguna 1975)''~\cite{LagunaPRB1975} respectively,
which were not produced by mechanical forcing (as in all other configurations) but by thermal forcing.
This thermal forcing sustains a velocity difference at large scales between the normal fluid and superfluid components, known as counterflow, and fuels a specific type of He-II turbulence. Interestingly, under such conditions, the prefactor appears to exhibit a peak around 1.9\,K that is not found in mechanically driven turbulence.

Finally, let us mention that by considering \eqref{eq:kolmogorov_scale} one predicts that
\begin{equation}
    \frac{\ell}{\etatilde}\approx 
    \frac{C_\omega^{1/4}}{2^{1/4}} 
    \left(\frac{\frac{\mu}{\rho}}{\kappa} \right)^{-1/2} \approx 2.7
    \label{eq:ell_over_eta}
\end{equation}
by assuming that $\nu_\text{tot} \approx \mu/\rho$, which is justified by the
conditions $\epsMF \ll \epsvisc$ (verified in Fig.~\ref{fig:energy_fluxes}) and $\rhos \nus \ll \mu$.
Therefore, the inter-vortex separation is found of the order of the Kolmogorov's microscale (in the temperature range of validity of the \HVBK{} modeling)
as originally argued by Vinen and Niemela~\cite{Vinen2002}. 
Remarkably, as shown in Table~\ref{tab:DNS_parameters}, the prediction \eqref{eq:ell_over_eta}
consistently agrees with the values of $\ell$
estimated independently from our numerical simulations (see Materials and Methods).}
Finally, it is worth noting that, since the Kolmogorov's microscale can be viewed as the small-scale cutoff of turbulent dynamics, the relation  $\ell/\tilde \eta = {\cal O}(1)$ aligns with the assumption that turbulent coarse-grained HVBK dynamics applies at scales much larger than $\ell$.

\begin{figure}[t]
  \centering
  \includegraphics[width = 0.95\linewidth]{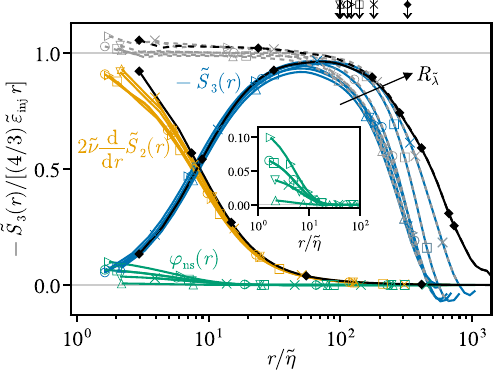}
  \caption{%
    Scale-by-scale energy budget of He-II turbulence.
    Colored curves represent the right-hand-side terms of \eqref{eq:four_thirds_law} at different temperatures and Reynolds numbers (symbols as in Table~\ref{tab:DNS_parameters}).
    All terms are compensated by ${4}/{3}\, \epsinj r$.
    The scale is normalized by the effective Kolmogorov scale defined in \eqref{eq:kolmogorov_scale}.
    Small downward arrows at the top of the figure indicate the forcing scale $\Lforcing$ for the different cases.
    For each case, the sum of the contributions (dashed gray lines) gives approximately one
    for increments over separation $r \ll \Lforcing$ indicating that \eqref{eq:four_thirds_law} is well satisfied.
    Inset:~details on the friction term $\phiNS(r)$ at small scales.
    }
  \label{fig:fourthirds}
\end{figure}

\section*{Scale-by-scale energy budget and intermittency}

The derived scale-by-scale energy budget \eqref{eq:KH_NS} is now examined through the simulations summarized in Table~\ref{tab:DNS_parameters}.
For velocity increments at scales $r$ smaller than the energy injection scale, the validity of \eqref{eq:four_thirds_law} is established in Fig.~\ref{fig:fourthirds}.  
At small scales, the third-order structure functions $\Stilde_3(r)$ (blue curves) collapse for the different cases, highlighting the relevance of the effective Kolmogorov scale \eqref{eq:kolmogorov_scale} % -- used to normalize the scale $r$ -- 
to characterize the small-scale behavior.
It is interesting to note that 
$\SSStilde(r)$ is also superimposed with its analogue for Navier-Stokes (NS) turbulence (in black). This implies that when considering the dynamics of the He-II mixture, the combined effects of viscous and frictional dissipation can be conceptualized as a global viscosity, $\nutot = \nutilde~\epsinj/\epsvisc$. This agrees with our previous definition of characteristic scales. 
Furthermore, as shown in Fig.~\ref{fig:energy_fluxes}, the mutual-friction dissipation term $\phiNS(r)$ (green curves) is more important at intermediate temperatures (see also inset of Fig.~\ref{fig:fourthirds}).
Nevertheless, even under these conditions, its contribution remains relatively modest compared with the viscous dissipation of the normal fluid.
Finally, at a fixed temperature and varying Reynolds numbers, the normalized curves for $\phiNS(r)$ overlap perfectly by keeping the ratio $\nu_n/\nu_s$ constant,  suggesting that this picture will not change notably at higher Reynolds numbers. However, $\phiNS(r)$ is enhanced (at fixed temperature and Reynolds number) if the kinematic viscosity $\nus$ attributed to the superfluid is reduced,
which follows the idea that mutual friction essentially compensates for the viscosity mismatch between the two fluid components, 
i.e. if the ``superfluid viscosity'' is lower, this is compensated by a greater dissipation by mutual friction.  

\begin{figure}[t]
  \centering
  \includegraphics[width= 0.95\linewidth]{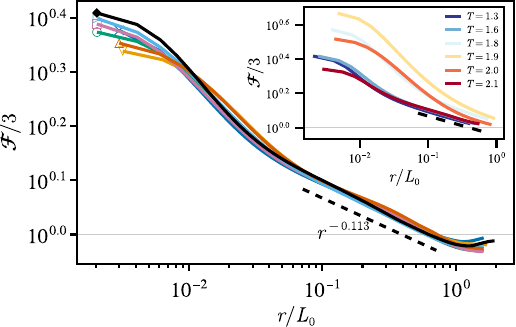}
  \caption{%
    Flatness of longitudinal velocity increments,
    $\Flatness(r) = \Emean{\dvpara(r)^4} / \Emean{\dvpara(r)^2}^2$,
    associated to the superfluid velocity $\vs$ (see Table~\ref{tab:DNS_parameters} for the temperatures).
    Inset: data from Biferale et al.~\cite{Biferale2018} at different temperatures.
    In the main panel and the inset, the black dashed lines represent the She--Lévêque prediction
    $\Flatness(r) \sim r^{-0.113}$ for classical turbulence~\cite{She1994}.
  }
  \label{fig:flatness}
\end{figure}

\subsection*{Intermittency}
As originally suggested by Batchelor~\cite{Batchelor1949},
a quantitative measure of intermittency is provided by 
the power-law scaling of the flatness of longitudinal velocity increments, $\Flatness(r)$, as the separation scale $r$ vanishes. Namely, 
\begin{equation}
    \mu(r)=-\frac{\mathrm{d} \log {\Flatness(r)}/{3} }{\mathrm{d} \log {r}/{L_0}}
    \label{def:intermittency_coeff}
\end{equation}
with $\Flatness(r) = \Emean{\dvpara(r)^4} / \Emean{\dvpara(r)^2}^2$. Let us recall that the normalization factor $3$ in \eqref{def:intermittency_coeff} corresponds to Gaussian statistics expected at scales $r \approx L_0$ of the order of the integral scale.  

The flatness $\Flatness(r)$ is displayed in Fig.~\ref{fig:flatness} for the
superfluid component
at different temperatures and Reynolds numbers (see Table~\ref{tab:DNS_parameters}).
The flatness obtained from classical turbulence data is also shown (black line). 
Two distinct behaviors are observed. A constant level of intermittency corresponding to the power-law exponent $\mu(r)\approx 0.1$ in a range of scales that can be associated with cascade dynamics, and an increase of intermittency attributed to the intensification of dissipative effects at smaller scales. 
The normal fluid displays similar features with the same intermittency exponent in the inertial range (see Fig. S1 in the Supporting information~\footnote{Supporting information can be found at \url{https://www.pnas.org/lookup/suppl/doi:10.1073/pnas.2426598122/-/DCSupplemental}}.).
The overall picture is very similar to that of classical turbulence \cite{Chevillard2005b}. It is important to note that the level of intermittency associated with dissipative dynamics changes with temperature and Reynolds number. In contrast, the level of intermittency in the cascade range exhibits a more robust behavior identical to that of classical turbulence. 
For comparison, the numerical results from \cite{Biferale2018} are displayed in the inset of Fig.~\ref{fig:flatness}.
Let us mention that in \cite{Biferale2018}, the flatness is plotted against the structure functions $S_3^{\sigma}(r)$ with $\sigma=(\mathrm{n},\,\mathrm{s})$ rather than against the length scale $r$.
It appears quite clearly that the enhancement of intermittency in the temperature range $1.8\,\mathrm{K} < T < 2.0\,\mathrm{K}$ is linked to dissipative rather than inertial dynamics.
At these temperatures, for which $\rhon \approx \rhos$, the frictional dissipation is maximal and affects all scales in these simulations, highlighting a smaller Reynolds number. 
Therefore, the controversy surrounding enhanced intermittency \cite{Biferale2018} appears to stem from a confusion between the considerations of inertial and dissipative scales, with the temperature dependence erroneously attributed instead of a finite Reynolds number effect.
Regarding the differences observed in shell-model simulations \cite{Boue2013,Shukla2016}, without providing explicit evidence here, we can mention that shell models, in which the coupling between modes is drastically decimated, are known to exhibit amplified bottleneck effects that depend on the Reynolds number and can affect inertial scalings  \cite{Kadanoff1995}.

The only experiment reporting a temperature-dependent intermittency effect \cite{Varga2018a} was carried out at a much lower Reynolds number compared to other experiments reporting intermittency measurements. Additionally, it featured a spatial resolution resolving the smallest flow scales of the normal fluid.  In this case, we may also suspect a certain confusion between the consideration of inertial and dissipative scales.

\section*{Discussion}

Our approach differs from previous studies by focusing on the behavior of the He-II mixture rather than its individual components. 
By extending the celebrated K\'arm\'an-Howarth equation to the \HVBK{} dynamics,
we establish a robust framework that allows us to extend the definitions of the Kolmogorov's and Taylor's microscales (and corresponding Reynolds numbers) to the two-fluid He-II mixture. 
A unified physical picture of He-II turbulence emerges, in which inertial scaling properties are similar to those of classical turbulence regardless of temperature, whereas temperature-dependent departures occur in the dissipative range. The dissipation associated with mutual friction is found to be significantly smaller than viscous dissipation in the explored range of temperatures.
Interestingly, in the temperature range where the two fluid components coexist (and \HVBK{} modeling is relevant), the inter-vortex distance is found to be of the order of the Kolmogorov's microscale, underscoring the regularity of the normal fluid flow on the length scales relevant to the interaction between quantized vortices. 

An important point concerning our work pertains to the influence of the \HVBK{} framework on the current results. This aspect fuels some ongoing debates \cite{Nemirovskii2019a, Nemirovskii2020, Nemirovskii2021}. In the present context of co-flowing isotropic He-II turbulence, it can be argued that the \HVBK{} framework offers a self-consistent approach to address friction without relying on adjustable ad-hoc parameters or external assumptions. 
Our study demonstrates its ability to shed light on the controversy regarding the alleged temperature dependence of intermittency, and to reconcile numerical simulations with experimental studies at high Reynolds numbers \cite{Rusaouen2017}. 
The \HVBK{} framework also allows us to derive an explicit law, 
${\ell}/{L_0} \approx 0.5 \,\mathrm{Re}_\kappa^{-3/4}$, which links the quantum and hydrodynamic features of He-II turbulent flows, and is found in excellent agreement with numerous experimental results.
Interestingly, the 0.5 prefactor can be reformulated as $\left({\nu_{\text{eff}}}/{(\kappa C_\epsilon)} \right)^{1/4}$ with the effective viscosity being estimated as $\nu_{\text{eff}} = \frac{C_\omega}{2} \frac{\mu}{\rho}$ rather than being twice this value, as is often assumed to match the kinematic viscosity of the classical case.

\section*{Materials and Methods}

\subsection*{Direct numerical simulations of \HVBK{} dynamics}

Numerical simulations of the \HVBK{} equations (\eqref{eq:HVBK}) with the friction force defined by~\eqref{eq:Fns_hvbk} were carried out.
The dynamics are integrated in a cubic domain of size $2\pi$ with periodic boundary conditions in all three directions, following previous studies \cite{Roche2009,Salort2011,Salort2012}.
Below, we summarize the key ingredients of the numerical method employed, which is standard in direct simulations of both classical and quantum isotopic turbulence.
We use a Fourier pseudo-spectral method~\cite{Canuto1988}, where the solution fields are represented as truncated Fourier series with $N^3$ degrees of freedom (see Table~\ref{tab:DNS_parameters} for values of $N$).
The spatial resolution is characterized by the maximum resolved wavenumber $\kmax = N / 2$.
As detailed in Table~\ref{tab:DNS_parameters}, we ensure $\kmax \etatilde > 1$ in all simulations, guaranteeing that the smallest turbulent structures (typically of size $\etatilde$) are properly resolved.
Nonlinear terms are computed in physical space using fast Fourier transforms and then transformed back into Fourier space, where time integration is performed using an explicit \emph{slaved} second-order Adams-Bashforth scheme. In this scheme, the viscous term is solved analytically and integrated directly in the scheme coefficients.
A constant timestep $\Delta t$ is chosen so that the CFL condition $v_{\text{max}} \Delta t / \Delta x \lesssim 1$ is satisfied at all times, where $v_{\text{max}}$ is the maximum velocity across the grid and $\Delta x = 2\pi/N$ is the grid cell size.
The two components are treated as incompressible, enforcing the conditions \ $\nabla \cdot \vn =0$ and $\nabla \cdot \vs = 0$, which determine the pressure fields.
Finally, to maintain a statistically stationary state, both components are isotropically forced at large scales within the shell of wavenumbers $1.5 \leq |\kvec| < 2.5$.  Identical energy injection rates are imposed, i.e.~$\epsinjn = \epsinjs$, following standard practice in classical turbulence studies \cite{Lamorgese2005}.
This results in a characteristic forcing scale $\Lforcing \sim 0.5$ for both components.

We consider three different temperatures, namely $T \approx 1.44$\,K (low), $T \approx 1.96$\,K (intermediate) and $T \approx 2.157$\,K (high), corresponding to the exact density ratios $\rhos / \rhon = 10$, $1$ and $0.1$ respectively.
The corresponding mutual-friction coefficient $B$ is set to $B=2.0$, $1.0$ and $2.16$ according to~\cite{Donnelly1998}.
The ``effective superfluid viscosity'' is fixed so that $\nus / \nun =1/5$ for all cases.
Additional cases with $\nus/\nun = 1/ 10$ have also been performed at $T \approx 1.96$K and $1.44$\,K to check possible dependence on  $\nus/\nun$.
For comparison, a classical turbulence dataset obtained from incompressible Navier-Stokes simulations~\cite{Polanco2021} is also included. 

\subsection*{Estimation of characteristic inter-vortex distance from numerical simulations}

To estimate the values of $\ell\equiv  1 / \Emean{\VLD}^{1/2}$ reported in
Table~\ref{tab:DNS_parameters}, the HVBK hypothesis
$|\vortS| = \kappa \VLD$ and the relation $\Emean{\VLD^2} = 2 \Emean{\VLD}^2$ \cite{Roche2008} are used to obtain $\ell^2 = \kappa / \OmegaS^{1/2}$, where $\OmegaS = \Emean{|\vortS|^2}/2$ is the superfluid enstrophy (readily estimated from the associated superfluid vorticity field).
Since simulations are performed in arbitrary units, the physical value of quantum of circulation, $\kappa \approx \SI{0.997e-7}{m^2/s}$, cannot be used directly in the above expression.
Instead, we invoke a similarity condition. Namely, that the dimensionless ratio $\gamma = \rhon \kappa / \mu $ --- computed using physical values of $\rhon$, $\kappa$ and $\mu$ --- must match $\kappa /\nun$ with values expressed in simulation units. From tabulated values of $\mu / \rhon$ at different temperatures~\cite{Donnelly1998}, we evaluate $\gamma$ and thereby deduce the respective values of $\kappa = \gamma \nun$ in simulation units. Specifically, we obtain $\gamma \approx 0.91$ (at 1.44\,K), $\gamma \approx 5.10$ (at 1.96\,K) and $\gamma \approx 5.87$ (at 2.157\,K).
% Values of \mu / \rhon = \nun obtained via https://www.mas.ncl.ac.uk/helium/ (based on Donnelly & Barenghi 1998)

\subsection*{The scale-by-scale energy budget of HVBK dynamics in the limit $r\to 0$}
In the scale-by-scale energy budget (\eqref{eq:four_thirds_law})
the contribution of the dissipation by mutual friction writes
\[
\phiNS(r) = A \, \kappa \Emean{\VLD} \, \frac{4}{r^2} \int_{0}^{r} \langle \vns(\xvec + \rvec')\cdot \vns(\xvec)\rangle  \, r'^2 \, \dd r'.
\]
As $r \to 0$, $\phiNS(r) \approx {4}/3 \, A 
\kappa \Emean{\VLD} \Emean{|\vns|^2} r$  at leading order.
Since $\pi r^2 \phiNS(r)$ represents the mean dissipation in a ball of radius $r$, one obtains that \[\phiNS(r) \approx \frac{4}{3} \epsMF r\]
where $\epsMF = A \kappa \Emean{\VLD}\Emean{|\vns|^2}$ may be interpreted as the mean frictional dissipation (per unit mass). 
The contribution of the viscous dissipation to the energy budget is given by the term
$ 2\, \nutilde \,{d\Stilde_2(r)}/{dr}$ with 
$\Stilde_2(r) \equiv  (\rhon \nu_n S_2^\mathrm{n}(r) + \rhos \nu_s S_2^\mathrm{s}(r))/\rho\nutilde$. \\
As $r \to 0$,  $\Stilde_2(r)\approx (\rhon\epsN  + \rhos \epsS)\, r^2 /\, 3\rho\nutilde = \epsvisc \, r^2 /\, 3\nutilde$ at leading order, where $\epsvisc$ corresponds to the viscous dissipation (per unit mass). This yields
\[ 2 \, \nutilde \,\frac{d\Stilde_2(r)}{dr} \approx \frac{4}{3} \epsvisc r.\]
Finally, the scale-by-scale budget (\eqref{eq:four_thirds_law})  reduces in the limit $r\to0$ to
\[\epsinj = \epsvisc + \epsMF\]
reflecting that the energy supplied to the fluid is dissipated though viscosity and mutual friction in He-II turbulence. Note that the term $\SSStilde(r) = {O}(r^3)$ does not contribute to this budget as $r\to 0$.

\begin{acknowledgments}
The authors acknowledge financial support from the French ANR grant ANR-18-CE46-0013 QUTE-HPC.
\end{acknowledgments}

\bibliography{Superfluids}

%IMPORTANT QUESTION IS HOW DOES ALL THIS DEPEND ON THE \HVBK{} MODEL ?
\end{document}